\documentclass[11pt,twoside]{article}

\usepackage{asp2004}
\usepackage{epsf}
\usepackage{psfig}
\usepackage{lscape}

\begin{document}
\title{Bayesian Model Selection and Extrasolar Planet Detection}   
\author{Eric B.\ Ford}
\affil{Hubble Fellow, Harvard-Smithsonian Center for Astrophysics, Mail Stop 51, 60 Garden Street, Cambridge, MA 02138, USA}
\author{Philip C. Gregory}
\affil{Physics and Astronomy Department, University of British Columbia, Vancouver, BC, V6T 1Z1, Canada}
\begin{abstract}
The discovery of nearly 200 extrasolar planets during the last decade
has revitalized scientific interest in the physics of planet
formation and ushered in a new era for astronomy.  Astronomers searching
for the small signals induced by planets inevitably face significant
statistical challenges.  For example, radial velocity (RV) planet
searches (that have discovered most of the known planets) are
increasingly finding planets with small velocity amplitudes, with long
orbital periods, or in multiple planet systems.  
Bayesian inference has the potential to improve the
interpretation of existing observations, the planning of
future observations and ultimately inferences concerning the overall population of planets. 
The main obstacle to applying Bayesian inference to extrasolar planet searches
is the need to develop computationally efficient algorithms for
calculating integrals over high-dimensional parameter spaces.  In
recent years, the refinement of Markov chain Monte Carlo (MCMC)
algorithms has made it practical to accurately characterize orbital
parameters and their uncertainties from RV observations of
single-planet and weakly interacting multiple-planet systems.  

Unfortunately, MCMC is not sufficient for Bayesian model selection,
i.e., comparing the marginal posterior probability of models with
different parameters, as is necessary to determine how strongly the
observational data favor a model with $n+1$ planets over a model with
just $n$ planets.  Many of the obvious estimators for the marginal
posterior probability suffer from poor convergence properties.  We
compare several estimators of the marginal likelihood and feature
those that display desirable convergence properties based on the analysis of a
sample data set for HD 88133b \citet{Fischer05}.  We find that methods
based on importance sampling are most efficient, provided that a good
analytic approximation of the posterior probability distribution is
available.  We present a simple algorithm for using a sample from the
posterior to construct a mixture distribution that approximates the
posterior and can be used for importance sampling and Bayesian model
selection.  We conclude with some suggestions for the development and
refinement of computationally efficient and robust estimators of
marginal posterior probabilities.
\end{abstract}


\section{Introduction}

Recent collaboration between astronomers and statisticians has led to a
better understanding of the particular challenges associated with
Bayesian analysis of dynamical planet detections.  In this paper, we
briefly review the state of the art of Bayesian parameter estimation,
model selection, and experimental design in the context of extrasolar
planet searches.  Then, we discuss recent work on Bayesian model
selection, demonstrating the properties of several estimators of the
marginal posterior probability using an actual set of data from an
RV planet search.

\subsection{Observational Data}

In RV surveys, the velocity of the central star is precisely monitored
for periodic variations which could be caused by orbiting companions
(see Fig.\ 1, left).  Each individual observation can be reduced to an
estimate of the observational uncertainty ($\sigma_{k}$) and a
measurement of the star's RV ($v_k$) relative to the $j_k$th velocity
reference at time $t_k$.
Because each RV measurement is based on calculating the centroid of
thousands of spectra lines and averaged over hundreds of sections of
the spectrum, the observational uncertainties of most current echelle
based RV surveys can be accurately estimated and are nearly Gaussian
\citep{Butler96}.  There may also be intrinsic stellar variability
(``jitter'') that we model as an addition source of uncorrelated
Gaussian noise with variance $s^2$ and add to the measurement
uncertainties in quadrature.
If the velocity observations ($\vec{v}_{\vec{\theta}}$) were generated
by the model specified by $\mathcal{M}$ and model parameters
$\vec{\theta}$, then the probability of measuring the observed
velocities is
\begin{equation}
L(\vec{\theta}) \equiv p(\vec{v}|\vec{\theta},  \mathcal{M}) = \prod_{k} \frac{1}{\sqrt{2\pi\left(\sigma_k^2+s^2\right)}} \exp \left[ -\frac{\left(v_{\vec{\theta}}(t_k,j_k)-v_k\right)^2}{2\left(\sigma_k^2+s^2\right)} \right],
\label{EqnLikelihood}
\end{equation}
assuming that the errors in individual observations and the ``jitter''
are both normally distributed and uncorrelated.

\begin{figure}[!ht]
\plottwo{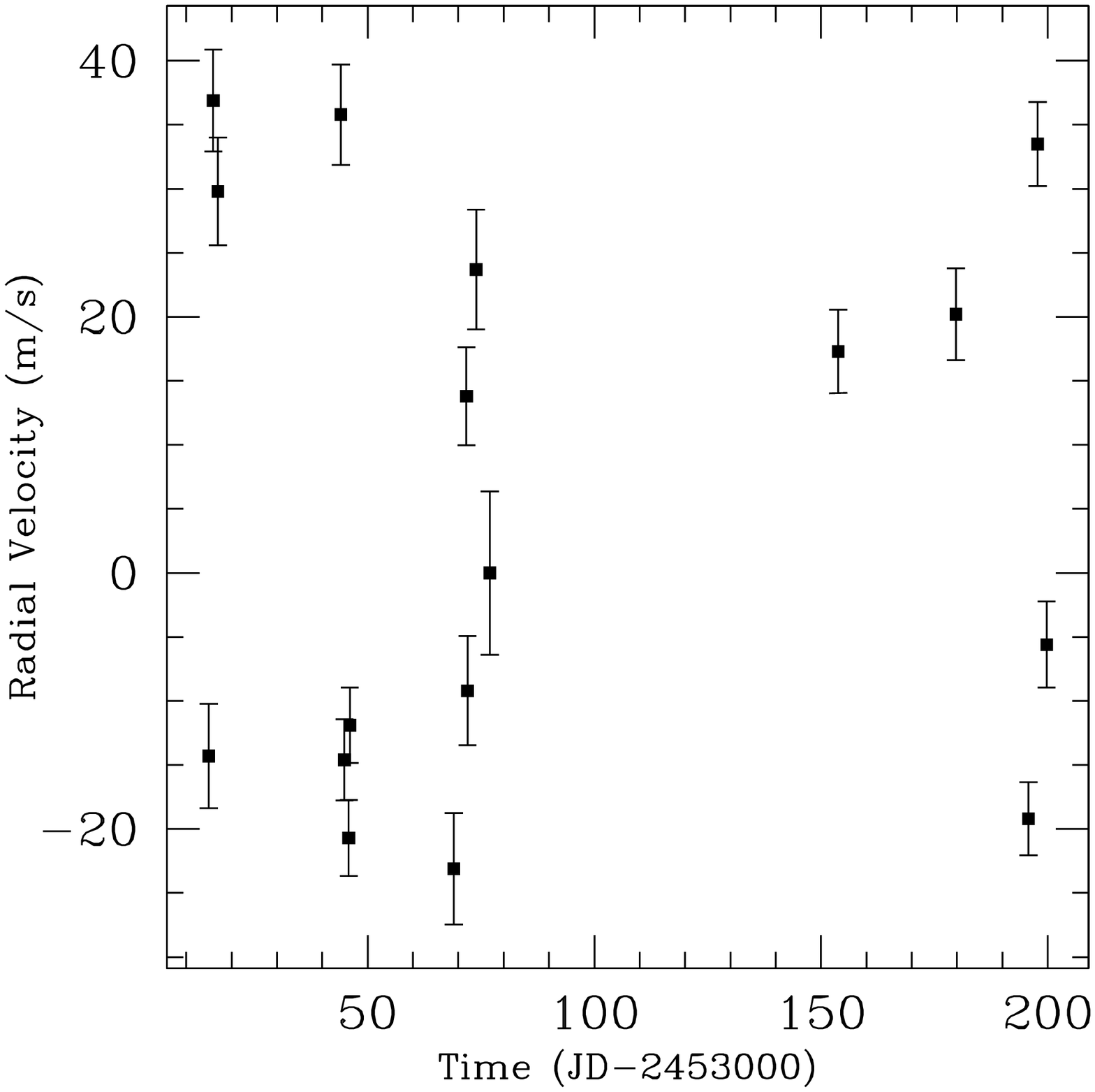}{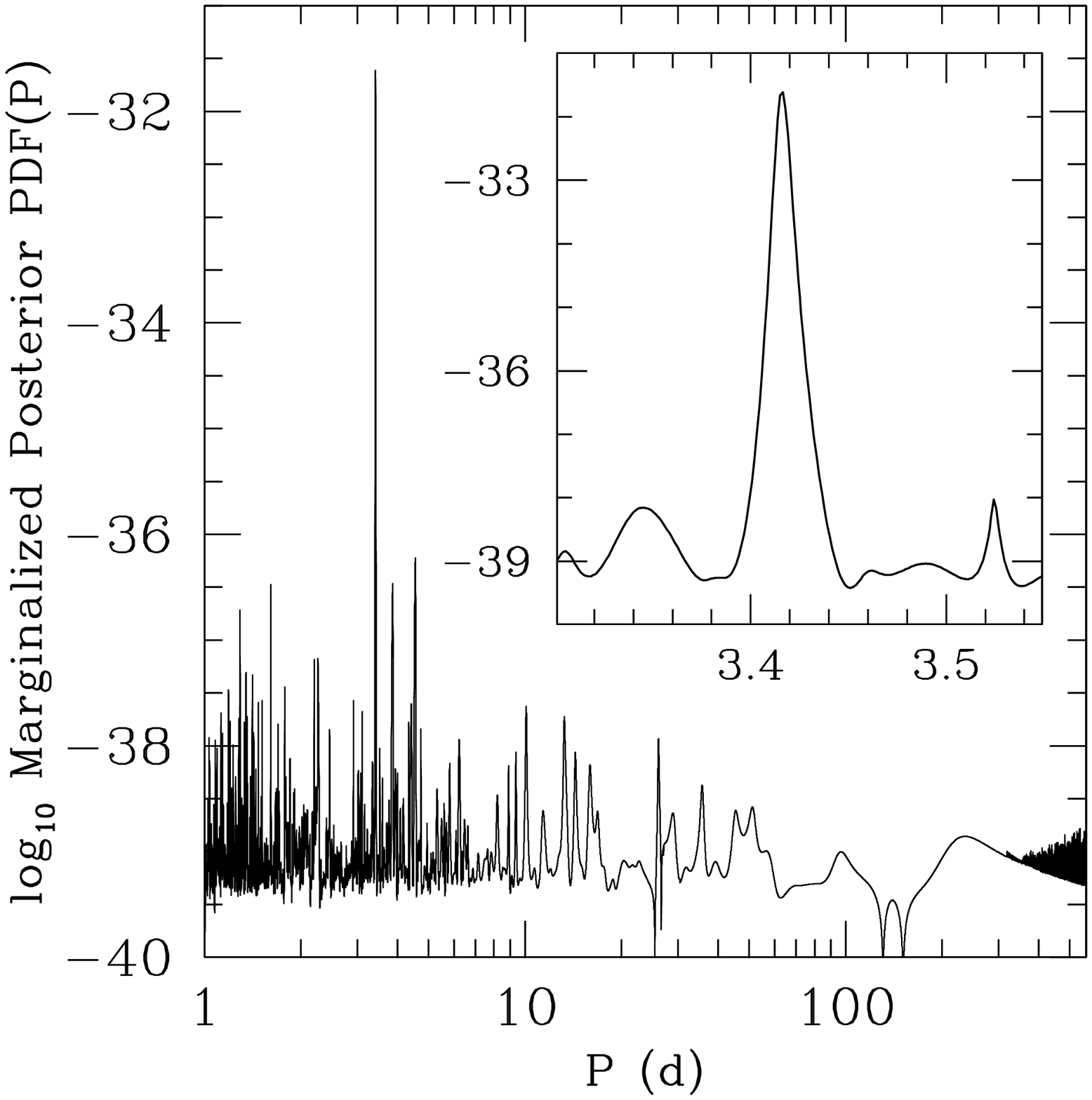}
\label{FigData}
\caption{Left: The 17 RV observations of HD 88133 published in
\citet{Fischer05} plotted versus the time of observation.  Right: The
log posterior probability marginalized over all model parameters
except the orbital period, assuming a single planet on a circular
orbit (i.e., the Bayesian generalization of the periodogram) and the
observational data shown on the left.}
\end{figure}

\subsection{Theoretical Model}

Specifying the mass and six phase space coordinates of each body in a
planetary system at a specified time provides a complete description
of the system.
In practice, it is convenient to choose the osculating Keplerian
orbital elements (orbital period, $P$, orbital eccentricity, $e$,
inclination relative to the plane of the sky, $i$, argument of
periastron measured from the plane of the sky, $\omega$, longitude of
ascending node, $\Omega$, and mean anomaly, $M$) for each planet in
Jacobi coordinates, since the mean anomaly is the only one of these
orbital elements that changes with time for a planet on an unperturbed
Keplerian orbit. The observed stellar velocity is the sum of the line
of sight velocity of the center-of-mass and the projection of the
reflex velocity due to any planetary companions onto the line of
sight.  For multiple planet systems, it can be important to use
complete n-body simulations to model the planetary motions accurately
(e.g., GJ876; Laughlin et al.\ 2005; Rivera et al.\ 2005).  However,
in most cases, the mutual planetary perturbations are negligible on
time scales comparable to the duration of observations.  In such
cases, the RV perturbations due to a multiple planet system can be
modeled as the linear superposition of multiple non-interacting
Keplerian orbits,
$
%
v_{\vec{\theta}}(t,j) = C_j + \sum_p \Delta v_{p}(t),
$
%
where $C_j$ is the $j$th velocity reference.  While there is a single
mean line of sight velocity of the center of motion, it is important
to use separate constants, $C_j$ for each observatory/spectrograph
pair, since the velocities are measured differentially relative to a
reference velocity that is unique to each observatory.  The
perturbation to the stellar RV ($\Delta v_{p}$) due to a planet on a
Keplerian orbit is given by
%
$
\label{rveqn}
\Delta v_{p}(t) = K_p \left[ \cos\left(\omega_p+T_p\right) + e_p \cos(\omega_p) \right] 
$
%
where $p$ labels the planet, $K$ is the velocity semi-amplitude and
$T$ is the true anomaly, which implicitly depends on time.  The true
anomaly ($T$) is related to the eccentric anomaly ($E$) via the
relation
%
$
\tan \left( T/2 \right) = (1+e)^{1/2}(1-e)^{-1/2} \tan^{1/2} \left( E/2 \right).
\label{EccTrueEqn}
$
%
The eccentric anomaly is related to the mean anomaly ($M$) via
Kepler's equation
%
$
\label{KeplerEqn}
E(t) - e \sin\left(E(t)\right) = M(t) - M_o = 2\pi t /P - M_o
$
%
where $M_o$ is a constant, the orbital phase at $t=0$.
Unfortunately, RV observations alone are not sensitive to the orbital
inclination relative to the plane of the sky ($i$) or the longitude of
ascending node ($\Omega$).  Therefore, RV observations by themselves
can only measure the minimum mass ratio, $m_{\min}/M_* = m \sin
i/M_*$, that is a function of $K$, $P$, $e$, and $M_*$, where $M_*$ is
the stellar mass, typically estimated by independent astronomical
observations.

\subsection{Bayesian Framework}

To quantitatively analyze the available observational constraints, we
employ the techniques of Bayesian inference.  By treating both the
observation and the model parameters as random variables, Bayesian
inference is able to address statistical questions in a mathematically
rigorous fashion.  The joint probability, $p(\vec{v}, \vec{\theta} |
\mathcal{M})$, can be expressed as the product of the likelihood
($L(\vec{\theta}) \equiv p(\vec{v} | \vec{\theta}, \mathcal{M})$), the
probability of the observables given the model parameters), and a
prior probability distribution function ($p(\vec{\theta} |
\mathcal{M})$) which is based on previous knowledge of the model
parameters.  Note that each of the probability distribution functions
(PDFs) is conditioned on the assumption of a model, $\mathcal{M}$,
that includes the meaning of the model parameters, $\vec{\theta}$, and
their relationship to the observational data, $\vec{v}$.  Bayes's
theorem allows one to compute a posterior probability density
function, $p(\vec{\theta} | \vec{v}, \mathcal{M})$, which incorporates
the knowledge gained by the observations $\vec{v}$,
\begin{equation}
p(\vec{\theta}| \vec{v}, \mathcal{M} ) = \frac{ p( \vec{\theta} | \mathcal{M} ) p(\vec{v} | \vec{\theta}, \mathcal{M} ) }{ \int p( \vec{\theta} | \mathcal{M} ) p( \vec{v}| \vec{\theta}, \mathcal{M} ) \,d\vec{\theta} }.
\label{BayesEqn}
\end{equation}
This paper is particularly interested in the case when there are
multiple viable models (e.g., no planet model, one planet model, two
planet model, etc.) for the current data set.  In this case, the
posterior for a given model and set of parameters is given by
$p(\vec{\theta}, \mathcal{M} | \vec{v} ) = p(\mathcal{M})
p(\mathcal{M}|\vec{v}) p(\vec{\theta} | \vec{v}, \mathcal{M} )$,
where $p(\mathcal{M})$ is the prior probability of model $\mathcal{M}$
and the marginal posterior probability for model $\mathcal{M}$ is
given by
\begin{equation}
\label{EqnMarginal}
m(\vec{v}) \equiv p(\mathcal{M} | \vec{v} ) =  \int p( \vec{\theta} | \mathcal{M} ) p( \vec{v}| \vec{\theta}, \mathcal{M} ) \,d\vec{\theta}.
\end{equation}
We briefly review recent progress in sampling from $p(\vec{\theta}|
\vec{v}, \mathcal{M} )$ via MCMC in \S1.5 and
introduce and compare algorithms for evaluating of $m(\vec{v})$ in \S2.

\subsection{Choice of Priors}

In Table~\ref{TabPriors}, we list the priors used for each model
parameter in this work.  While the choice of these and other
parameters can be fine-tuned for a given problem, we suggest these
choices as a starting point for the Bayesian analysis of radial
velocity data sets.  Physical and geometric considerations lead to
natural choices for the prior PDFs for most of the model parameters
(see Table~\ref{TabPriors}).  A few of the priors merit closer
attention.  The cutoff at a minimum orbital period is chosen to be
less than the smallest orbital period of known extrasolar planets, but
this is somewhat larger than the theoretical limit (the Roche limit
occurs at $\sim0.2$d for a 10$M_{\mathrm{Jup}}$ planet around a
1$M_{\odot}$ star).  The cutoff at a maximum orbital period is much
longer than any known extrasolar planet, but is chosen to be roughly
where perturbations from passing stars and the galactic tide would
disrupt the planet's orbit.  The cutoff at a maximum velocity
semi-amplitude is chosen to be a function of orbital period, so that
the cutoff corresponds to a constant planet-star mass ratio.  In this
paper, we set $K_{\max} = 2129$m/s, which corresponds to a maximum
planet-star mass ratio of 0.01.  This choice is primarily based on the
observed distribution of extrasolar planet masses.  Clearly, there are
stellar binaries with much larger velocity amplitudes, but this limit
can be considered the definition of a planet.  While the possibility
of arbitrarily small masses (and hence velocity amplitudes) prevents a
physical justification for a lower cutoff, $K_{\min}$, it is not
possible to detect or constrain the orbital parameters of a
sufficiently low-mass planet.  To keep the prior distribution
normalizable, we impose breaks in the priors for $K$ and $s$ at $K_0$
and $s_0$, chosen to be at a velocity amplitude less than the smallest
detectable velocity amplitude.  For a data set with $N_{\mathrm{obs}}$
RV measurements with typical measurement precision of $\bar{\sigma}$,
we suggest setting $K_0 \le \bar{\sigma} \sqrt{50/N_{\mathrm{obs}}}$
(Cumming 1999).  For the reference prior, we chose $s_0 = K_0 = 1$m/s,
somewhat arbitrarily, but motivated by the current state of the art in
RV planet searches \citep{N2K6}.  For systems where a planet is
clearly detected, the shape of the posterior will not be sensitive to
our assumptions about the prior for small values of $K$, but for
planets which are marginally detected, this choice may become
significant. If the posterior distribution has significant probability
near $K\simeq K_o$, then one should check how sensitive any
conclusions are to the choice of $K_o$.

\begin{table}[!ht]
\caption{SAMSI Exoplanet Working Group Reference Priors}
\smallskip
\begin{center}
{\small
\begin{tabular}{ccccc}
\tableline
\noalign{\smallskip}
Model Parameter & Variable & Prior Distribution & Minimum & Maximum \\
\noalign{\smallskip}
\tableline
\noalign{\smallskip}
Amplitude of jitter & $s$ & $\frac{(s+s_0)^{-1}}{\log\left(1.+\frac{s_{\max}}{s_0}\right)}$ & 0 m/s & $K_{\max}$ \\
\smallskip\\
\multicolumn{5}{c}{Parameters for each velocity reference} \\
Velocity offset & $C_j$ & $\frac{1}{C_{\max}-C_{\min}}$ & -$K_{\max}$ & $K_{\max}$ \\
\smallskip\\
\multicolumn{5}{c}{Parameters for each planet} \\
Orbital period & $P_i$ & $\frac{P_i^{-1}}{\log(P_{\max}/P_{\min})}$ & 1 d & $10^3$ yr \\
Velocity semi-amplitude & $K_i$ & $\frac{(K_i+K_0)^{-1}}{\log\left[1.+\frac{K_{\max}}{K_0}\left(\frac{P_{\min}}{P_i}\right)^{1/3}\right]}$ & 0 m/s & $K_{\max}$ $\left(\frac{P_{\min}}{P_i}\right)^{\frac{1}{3}}$ \\
Orbital eccentricity & $e_i$ & 1  & 0 & 1 \\
Argument of periastron & $\omega_i$ & $\frac{1}{2\pi}$ & 0 & $2\pi$ \\
Orbital phase  & $M_i$ & $\frac{1}{2\pi}$ & 0 & $2\pi$ \\
\noalign{\smallskip}
\tableline
\end{tabular}
}
\end{center}
\label{TabPriors}
\end{table}

\subsection{Previous Research}

Identifying the correct orbital solution from a set of RV (or other
dynamical) observations is challenging due to the necessity of
considering a very large parameter space of possible solutions (see
Fig.\ 1, right).  For RV planet searches, there are at least five
model parameters per planet and one model parameter per observatory.
To make the global search problem tractable, RV data sets are
traditionally searched for sinusoidal signals (potential planets)
using a periodogram.  The advantage of the periodogram is that it is
extremely tractable computationally ($O(n\log n)$) and can be quite
useful for identifying potential orbital periods.  Then
Levenberg-Marquardt (LM) minimization \citep{NR} is applied starting
from several initial guesses of orbital solutions near each of the
potential orbital periods identified by the periodogram.  If the
quality of the fit is consistent with the combination of the
observational uncertainties \citep{Butler96} and the expected
intrinsic stellar variability \citep{Wright05}, then estimates of the
uncertainties in orbital parameters are obtained by repeatedly finding
the best-fit orbital parameters (with LM) to several synthetic data
sets generated via bootstrap resampling (with replacement) of the
observational data \citep{NR}.  These methods work quite well for
analyzing stars with a single planet on a low-eccentricity,
short-period (relative to the duration of the observations) orbit when
the velocity perturbation is large (relative to the measurement
uncertainties).  Since such planets are the easiest to discover, they
are common among the known sample of planets, and the traditional
frequentist methods have proven quite valuable in their discovery.

In recent years, RV searches have become increasingly sensitive to
planets with small velocity amplitudes and/or long orbital periods, as
well as planets in multiple planet systems.  Bayesian inference can
help with each of these challenges and has the potential to
significantly improve the sensitivity of detections and accuracy of
orbital determinations.  Recent work has begun to develop the
framework and computational tools to make this happen.  For example,
\citet{Cumming04} discussed the relationship between the periodogram
method and a Bayesian analysis that assumes any planet is on a
circular orbit.  \citet{Ford06Adapt} combined brute force Monte Carlo
(to integrate over orbital period) and the Laplace approximation (to
integrate over the remaining model parameters) to render Bayesian
model selection practical for planets assumed to be on a circular
orbit (e.g., short-period planets prone to tidal circularization).
MCMC has been applied to estimate orbital parameters and their
uncertainty and to help understand the situations where traditional
frequentist methods leave significant room for improvement
\citep{Ford05Mcmc,DriscolMTh}.  \citet{Gregory05Book} and
\citet{Ford06Fast} have automated the determination of parameters of
the candidate transition PDFs.  \citet{Ford06Fast} has also identified
non-linear candidate transition PDFs that dramatically accelerate the
convergence of MCMC.  These advances make it computationally feasible
for MCMC to characterize the allowed orbital solutions for single
planets or weakly-interacting multiple planet systems
\citep[e.g.][]{Ford05UpsAnd}.  \citet{Gregory05AAT} has applied
parallel tempering to allow MCMC to explore multiple orbital solutions
widely separated in parameter space.
\citep{Loredo04} developed the theoretical framework for applying
Bayesian adaptive experimental design to dynamical extrasolar planet
searches, and \citet{Ford06Adapt} developed computationally practical
algorithms for applying these techniques to adaptively schedule radial
velocity observations.  
Bayesian techniques are just beginning
to be applied to analyze the population of extrasolar planets
\citep{Ford06Tidal,Loredo06}.

\section{Algorithms for Applying Bayesian Model Selection To Extrasolar Planets}

While MCMC techniques have proven very efficient for sampling from the posterior
distribution for orbital parameters of extrasolar planets
\citep[][e.g.]{Ford06Fast}, MCMC does not directly determine the normalizing
constant of the posterior distribution.  While this is not necessary
for parameter estimation (within a single model), it is essential when
considering multiple possible models (e.g., no planet, one planet, two
planets...).  \citet{Clyde06} reviews the state of modern techniques
for Bayesian model selection from a statistics perspective.  In this
paper, we introduce several estimators of the marginal posterior and
test their performance on the radial velocities for HD 88133 published
in \citet{Fischer05}.  
While our test data set consists of purely RV
observations, we expect that most of our findings are also directly
applicable to other dynamical planet searches (e.g., astrometric,
pulsar/white dwarf timing).  Other types of planet searches (e.g.,
transits, microlensing, direct imaging) likely present different
challenges.  Nevertheless, the challenge of estimating marginal
posterior probabilities is quite general.  Thus, we expect our finding
may provide insights into methods for Bayesian model selection in
other areas of astronomy and statistics.

\subsection{Sampling from Prior}

\subsubsection{Basic Monte Carlo}

The most obvious basic Monte Carlo (BMC) estimator of $m(\vec{v})$ is
based on drawing $\vec{\theta}_i$ from the prior and discretizing the
integral in Eqn.\ \ref{EqnMarginal} to create the estimator
$
\hat{m}_{BMC}(\vec{v}) = \sum_{i=1}^{n} L(\vec{\theta}) / n
$.
Unfortunately, the very large
parameter space makes this totally impractical, even for a single
planet system.  Using the prior in Table~\ref{TabPriors}, we drew
over $10^9$ samples, but $\hat{m}_{BMC}(\vec{v})$ underestimated
$m(\vec{\theta})$ by orders of magnitude while the internal error
estimate suggested a random error of 2\%.  This is due to the
fact that not a single sample landed in the dominant peak in the
likelihood (see Fig.\ 1, right).  

\subsubsection{Restricted Monte Carlo}

The BMC estimator can be easily modified by sampling from only a small
subset of the prior.  Using MCMC, we sample from the posterior, select
one model parameter for investigation and then marginalize over all
the remaining parameters.  Then, we use the marginalized posterior
distributions to identify the subset of parameter space with
non-negligible probability (e.g., 99.9\% credible interval).  Using
this technique, we identified a region with volume, $V_{RMC}$,
$\simeq2\times10^8$ times less than the volume of the prior
distribution, $V_{\mathrm{Prior}}$.  Then, we can estimate,
\begin{equation}
\widehat{m}_{RMC}(\vec{v}) = \frac{V_{RMC}}{V_{\mathrm{Prior}}} \sum_{i=1}^{n} L(\vec{\theta_i}) / n,
\end{equation}
where $\vec{\theta}$ are drawn from a distribution proportional to the
prior over the restricted range of parameter values and zero
elsewhere.  For our test case, the RMC estimator provides a reasonable
estimate of $\hat{m(\vec{v})}_{RMC}$ (see Fig.\ 2, top, solid curve).  However, this
estimator has several short comings.  First, it is biased due to the
fact that it includes the probability coming from only one hypercube
of parameter space.  If we choose a large subvolume, then the
estimator converges slowly, since most samples miss the high
likelihood regions.  If we choose a small subvolume, then we may
neglect a significant region of probability outside our hypercube.
These problems are exacerbated for data sets where there are
significant correlations between parameters and/or many model
parameters.  While we were able to choose an effective subvolume for
the test data set, the estimator converged slowly, requiring
$\sim2\times10^6$ samples to reach 5\% accuracy.  Finally, we note
that prohibitively large subvolumes can be necessary for other data
sets that allow a broader range of orbital solutions and/or
significant correlations between parameters.  

\subsubsection{Partial Linearization \& Laplace Approximation}

In the approximation of circular orbits, the predicted velocity can be
written as a linear function of functions of all the model parameter
except $P$ and $s$.  Thus, for given values of $P$ and $s$, there is a
single global maximum of the likelihood that can be quickly located by
solving a set of linear equations.  We can analytically integrate over
the remaining model parameters ($K_i$, $M_i$, $C_j$) using the Laplace
approximation by evaluating the likelihood at the global maximum (for
a given $P$ and $s$) and the Fischer information matrix evaluated at
that point \citep[for details see][]{Cumming04,Ford06Adapt}.  For a
system with $N_p$ planets, this leaves only 1+$N_p$ dimensions to be
explored by brute force Monte Carlo, making the Monte Carlo
integration dramatically more efficient.  For small eccentricities,
the velocity perturbation due to a planet can be approximated by a
series expansion in eccentricity.  If we use the $O(e^1)$ (epicycle)
approximation, then the predicted velocity can again be linearized
over all model parameters except $P$ and $s$, allowing the this
technique to efficiently consider for small eccentricities.  We choose
a test case to have a modest velocity amplitude and small
eccentricity, so that the circular and epicycle approximations provide
reasonable approximations for the radial velocity perturbations due to
HD 88133b.  Thus, we expected partial linearization would provide at
least an order of magnitude estimate of the true marginal posterior
probability, as well as a point of reference for comparing other
estimators.  Indeed, a comparison of this estimator to the more
sophisticated estimators of \S2.3-2.5, we find that this estimator of
$m(\vec{v})$ shows a systematic bias (Fig.\ 2, top, long-dashed curve), likely due
to the Laplace and the epicycle approximations.  When the linear
approximation to the velocity is not adequate, then a similar
technique can be used, but the predicted velocity is non-linear in the
parameters $P_i$, $e_i$, $M_i$, and $s$, so the partial linearization
leaves $1+3N_p$ dimensions to be explored by brute force.  While
partial linearization is useful for analyzing systems with one planet
\citep[e.g.,][]{Ford06Adapt}, it rapidly becomes computationally
impractical for multiple planet systems.

\begin{figure}[!ht]
\plottwo{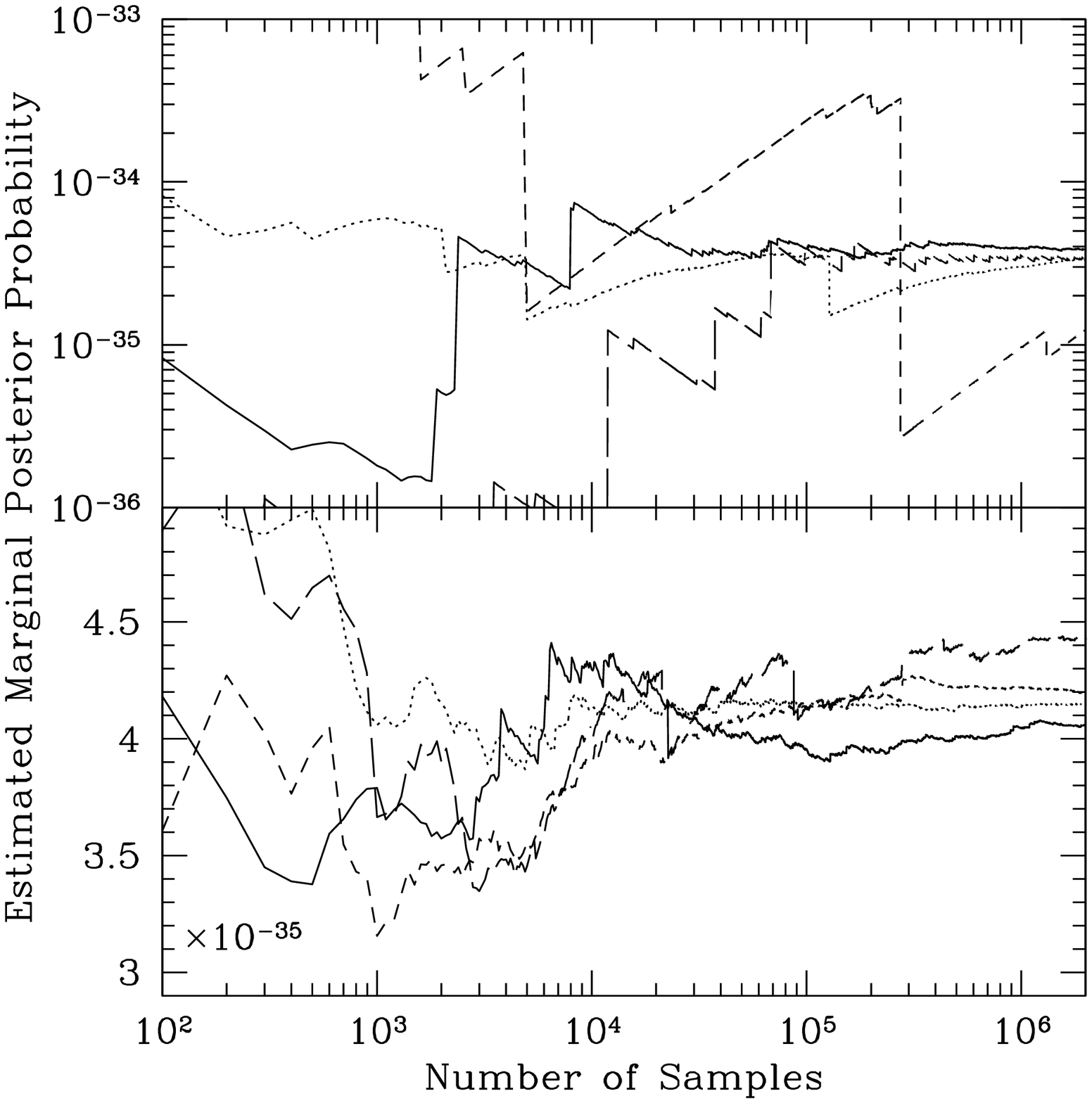}{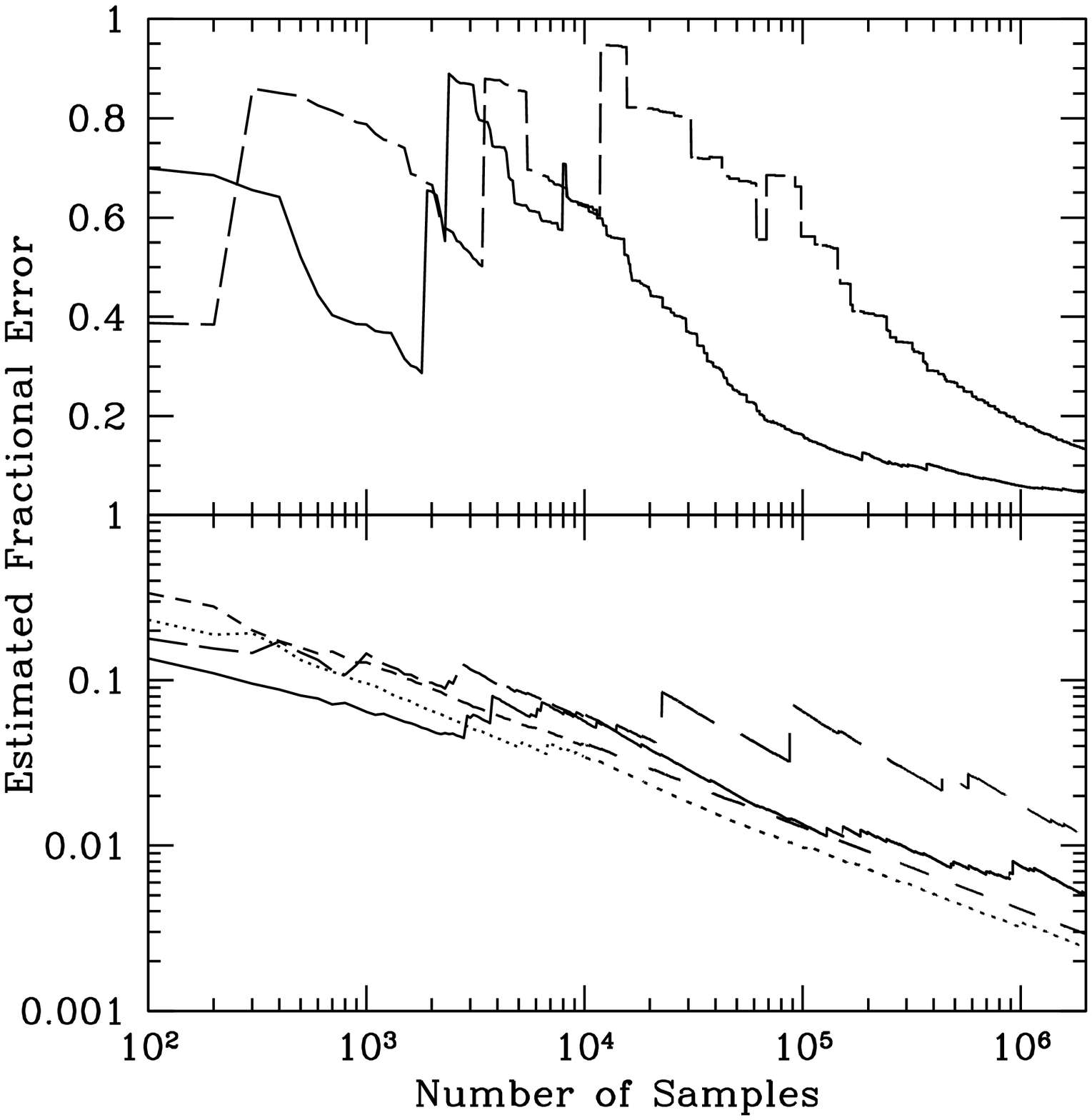}
\label{FigMarginalVsSamples}
\caption{Comparison of several estimators of the marginal posterior
probability for single planet models and the 17 RV
observations of HD88133 shown in Fig.\ 1.  Left: Estimators of the
marginal probability for a one-planet model.  Right: Internally
estimated standard deviation of each estimator.  The line styles indicate the
algorithms used for each estimator: Restricted Monte Carlo
(top, solid), Partial Linearization (top, long-dash),
Harmonic Mean (top, short-dash), Gelfand \& Dey with Partial
Linearization (top, dotted), Importance Sampling from
single Normal (bottom, dotted), Importance Sampling from Mixture of
Normals (bottom, solid), Gelfand \& Dey with Mixture of Normals
(bottom, long-dash), Ratio Estimator (bottom, short-dash).  }
\end{figure}

\subsection{Sampling from Posterior}
Given the serious difficulties in sampling from the prior, we proceed
to consider estimators of the marginal posterior that sample from
alternative distributions.  MCMC provides a
computationally efficient tool for sampling from the posterior.  Since
a Bayesian analysis will typically use MCMC to generate a sample from
the posterior for the purposes of parameter estimation, we investigate
estimators that can use such a sample to calculate $m(\vec{v})$.

\subsubsection{Harmonic Mean}
Newton \& Raftery (1994) propose the estimator
$
\hat{m}_{NR}(\vec{v}) = n / \sum_{i=1}^n 1/L(\vec{\theta_i})
$.
Unfortunately, this estimator displays extremely poor convergence
properties (Fig.\ 2, left, top, short-dash
curve), as it has an infinite variance.  Newton \& Raftery (1994)
suggested sampling from a mixture of the posterior and prior to obtain
an estimator with finite variance, but we found similarly poor
performance.

\subsubsection{Weighted Harmonic Mean \& Partial Linearization}
Gelfand \& Dey (1994) use the identity 
\begin{equation}
\left[m(\vec{v})\right]^{-1} = \int \frac{h(\vec{\theta})}{p(\vec{\theta})L(\vec{\theta})} p(\vec{\theta}|\vec{v}) d\vec{\theta},
\end{equation}
where $h(\vec{\theta})$ is an arbitrary density to create an estimator for the marginal posterior probability,
\begin{equation}
\widehat{m}_{WHM,h}(x) = n / \sum_{i=1}^{n} \frac{h(\vec{\theta}^{i})}{p(\vec{\theta}^{i}) L(\vec{\theta}^{i})},
\end{equation}
where $\theta^{i}$ are drawn from the posterior.  This estimator
should perform well when $h(\vec{\theta})$ approximates the posterior
in regions of high probability, and it has finite variance when the
tails of $h(\vec{\theta})$ decay faster than the tails of
$p(\vec{\theta}) L(\vec{\theta})$ \citep[see][\S6.3.1]{CarlinLouis00}.
We set $h(\vec{\theta})$ equal to $p_e(\vec{\theta} | \vec{v} ) =
p(\vec{\theta}) L_{e}(\vec{\theta}) / \int d\vec{\theta}\,
p(\vec{\theta}) L_{e}(\vec{\theta})$, where $L_{e}(\vec{\theta})$ is a
likelihood computed as in Eqn.\ \ref{EqnLikelihood}, but replacing the
Keplerian model with the epicycle approximation for computing the
velocity perturbation of each planet and $p_e(\vec{\theta} | \vec{v}
)$ is the posterior under this approximation.  The normalization is
calculated using partial linearization \& the Laplace approximation as
discussed in \S2.1.3.  Unfortunately, we find that even this
approximation displays poor convergence in our test case (Fig.\ 2,
left, top, dotted curve).

\subsection{Simple Importance Sampling}
Given the poor convergence properties of the previous estimators based
on sampling from the prior and posterior densities, we investigate
estimating $m(\vec{v})$ via importance sampling.  Importance sampling
requires that we specify a properly normalized density
$h(\vec{\theta})$ which we can both evaluate at any $\vec{\theta}$ and
sample from efficiently.  We estimate $m(\vec{\theta})$ with
\begin{equation}
\widehat{m}_{IS,h}(\vec{\theta}) = 
\frac{1}{n}\sum_{i=1}^n \frac{p(\vec{\theta}^i)L(\vec{\theta}^i)} {h(\vec{\theta}^i)},
\end{equation}
where $\vec{\theta}^1, \ldots,\vec{\theta}^n$ are drawn from $h(\vec{\theta})$.
We choose $h(\vec{\theta}) = N(\vec{\vartheta}_o, \Sigma)$, where $N$
is the multivariate normal distribution in a set of transformed model
parameters, $\vec{\vartheta}(\vec{\theta})$ \citep{Ford06Fast}.  We
use the transformation $\vec{\vartheta}(\vec{\theta}) = \left\{ \log P
\right.$, $K\cos\left(\omega+M\right)$, $K\sin\left(\omega+M\right)$,
$e\cos\omega$, $e\sin\omega$, $C$, $\left.\log s \right\}$ to help
reduce non-linear correlations between model parameters.  We use a
sample from the posterior to determine the location,
$\vec{\vartheta}_o$, and sample covariance, $\Sigma'$, in the
transformed coordinates.  We find that the variance of this estimator
is reduced if we scale the sample covariance matrix by a factor
$\varsigma\simeq2$ to obtain $\Sigma = \varsigma \Sigma'$, the
covariance matrix used in the importance sampling density.

In our test case, $\hat{m}_{IS,N}$ appears to be a very robust and
efficient estimator for this data set (Fig.\ 2, bottom, dotted curve).  Indeed, there would
be no need to pursue more sophisticated estimators, if
$\widehat{m}_{IS,h}(\vec{\theta})$ performed so well on all data sets.
However, we are concerned that this estimator will not be viable for
other data sets where the posterior can not be well approximated by a
single multivariate normal distribution.  This is likely to occur in
systems with long orbital periods \citep[see][Fig.\ 1]{Ford05Mcmc},
small data sets, and/or when the velocity amplitude is small.  The
posterior typically is dominated by a single peak for most published
RV data sets (almost by definition, since a data set that can be
explained by two qualitatively different orbital solutions would not
be considered to have discovered the planet and is unlikely to be
published).  However, if Bayesian model selection is to be used for
deciding when a planet has been detected or as a part of Bayesian
adaptive design, then it will be necessary to analyze data sets before
the posterior is so strongly peaked that it can be well approximated
by a single normal distribution.  Therefore, we proceed to develop a
more sophisticated importance sampler that can be more robust when
analyzing such data sets.

\subsection{Sampling from a Mixture Density}
\label{SecMixture}

\subsubsection{Basic Importance Sampling}
In an effort to develop an importance sampling density suitable for
application to a general RV data set, we consider a mixture of
multivariate normal distributions.  We assume that MCMC has already
been used to obtain a good sample ($\vec{\theta}^1, \ldots,
\vec{\theta}^{n_t}$) from the posterior and use this to construct an
importance sampling density,
\begin{equation}
g(\vec{\theta}) = \frac{1}{n_c} \sum_{j=1}^{n_c} g_j(\vec{\theta}),
\end{equation}
where we have randomly chosen $n_c=100$ samples to be removed from the original posterior sample
and to be used as the locations for the mixture components, $g_j(\vec{\theta})$.  
We choose 
each mixture
component to be a multivariate normal distribution, 
$
g_j(\vec{\theta}) = N(\vec{\vartheta}(\vec{\theta}) | \vec{\vartheta}(\vec{\theta}^j), \Sigma_j )
$,
where we must determine a covariance matrix for each $g_j$ using the
posterior sample.  First, we compute $\vec{\rho}$, defined to be a
vector of the sample standard deviations for each of the components of
$\vec{\vartheta}$, using the posterior sample.  Next, we define the
distance between the posterior sample $\vec{\theta}^i$ and the center
of $g_j(\vec{\theta})$,
$
d^2_{ij} = \sum_{k} \left(\vartheta_k(\vec{\theta}^i) - \vartheta_k(\vec{\theta}^j) \right)^2 /
\rho_k^2
$,
where $k$ indicates the element of $\vec{\vartheta}$ and $\vec{\rho}$.
We draw another random subset of $n_{cv}=50 n_c$ samples from the
original posterior sample (without replacement), select the
$n_{cv}/n_{m}$ posterior samples closest to each mixture component and
use them to calculate the covariance matrix, $\Sigma'_j$, for each
mixture component.  We set $\Sigma_j = \varsigma \Sigma'_j$, and
$\varsigma=1$.  Thus, we have developed an automated algorithm for
using a posterior sample to construct an importance sampling density,
$g(\vec{\theta})$.  Since the posterior sample is assumed to have
fully explored the posterior, $g(\vec{\theta})$ should be quite
similar to the posterior in all regions of significant probability,
provided that we use enough mixture components.

We use $g(\vec{\theta})$, $n_s$ samples from the remainder of our
posterior sample, and deterministic mixture sampling to compute the
estimator, $\widehat{m}_{IS,g}(\vec{v})$.  We find that it performs
quite well in our test case (Fig.\ 2, bottom,
solid curve).  It converges nearly as rapidly as
$\widehat{m}_{IS,N}(\vec{v})$ and appears somewhat more robust, even
for the rather simple posterior in our test case.  In tests on more
complex data sets, we find that $\widehat{m}_{IS,g}(\vec{v})$ can be
significantly more robust than $\widehat{m}_{IS,N}(\vec{v})$ for data
sets with somewhat less well constrained posterior PDFs, but both
estimators perform poorly on other data sets with very diffuse
posterior PDFs.  We speculate that our method for choosing the mixture
components could be replaced by a more sophisticated algorithm that
might result in a superior importance sampling densities for
challenging data sets.

\subsubsection{Defensive Importance Sampling}
Despite the success of $\widehat{m}_{IS,g}$, we have some concerns about
the robustness of $\widehat{m}_{IS,g}$ for high dimensional parameter
spaces (e.g., analyzing systems with several planets).  As the number
of model parameters increases, it will become increasingly difficult
to avoid $p(\vec{\theta}) L(\vec{\theta}) / g(\vec{\theta})$ becoming
unusually large for some values of $\vec{\theta}$.  To prevent this
we generalize our importance sampling density to include a component
from the prior, by defining $g_0(\vec{\theta}) = p(\vec{\theta})$ and
$g^*(\vec{\theta}) = \sum_{j=0}^{n_c} g_j(\vec{\theta}) / (n_c+1)$,
Following \citet{OwenZhou00Safe}, we combine this mixture density with
control variables to obtain the estimator
\begin{equation}
\widehat{m}_{DIS,g^*,\vec{\beta}}(\vec{v}) = \frac{1}{n_s} \sum_{i=1}^{n_s} \frac{p(\vec{\theta}^i) L(\vec{\theta}^i) - \sum_{j=0}^{n_c} \beta_j g_j(\vec{\theta}^i)}{g^*(\vec{\theta}^i)} + \sum_{j=0}^{n_c} \beta_j,
\end{equation}
which is valid for any choice of $\vec{\beta}$.  To minimize the
 variance of this estimator, we set $\vec{\beta}=\vec{\beta}^*$, where
$ \vec{\beta}^*$ is determined by least squares fitting to the linear
 system of $n_s$ equations
\begin{equation}
\sum_{j=0}^{n_c}  \left(\frac{g_j(\vec{\theta}^i)}{g^*(\vec{\theta}^i)}\right) \beta^*_j  = \frac{p(\vec{\theta}^i)L(\vec{\theta}^i)}{g^*(\vec{\theta}^i)}
\end{equation}
\citet{OwenZhou00Safe} show that this estimator is never worse than an
estimator based on any subset of the mixture components.  In practice,
we need $n_s$ to be large to have small variance, but it is not
practical to solve the linear system of equations with
$n_s\simeq10^{5-7}$.  Therefore, we repeatedly solve for $\vec{\beta}$
using subsets of $n_r \ll n_s$ posterior samples and average the
results to estimate $\vec{\beta}^*$.  The resulting estimator
$\widehat{m}_{DIS,g^*,\vec{\beta}^*}(\vec{v})$ is at least as good as
$\widehat{m}_{IS,g}(\vec{v})$ and is expected to be considerably more
robust.  In our test case, the estimators
$\widehat{m}_{IS,g}(\vec{v})$ and
$\widehat{m}_{DIS,g^*,\vec{\beta}^*}(\vec{v})$ follow each so closely
that the curves would be indistinguishable in Fig.\ 2 (bottom, solid
curve).  This is because the values of $\beta_j^*$ are roughly
comparable for all components with $j\ge1$, while $\beta_0$ (the prior
component) is orders of magnitude less than the other $\beta_j$.  This
reflects the fact that $g(\vec{\theta})$, our mixture of $n_c=100$
multivariate normal distributions, was sufficient to accurately
approximate the posterior density.  While
$\widehat{m}_{DIS,g^*,\vec{\beta}^*}(\vec{v})$ is somewhat more
computationally expensive, we still prefer it to
$\widehat{m}_{IS,g}(\vec{v})$, since it should be more robust.
Further, if $g(\vec{\theta})$ was inadequate, then $\beta_0^*$ would
increase, alerting us to the potential weakness of $g(\vec{\theta})$.

\subsubsection{Weighted Harmonic Mean}
We now reconsider the estimator of Gelfand \& Dey (1994), but using
the mixture, $g(\vec{\theta})$, for the weight function $h$.  The
resulting estimator $\widehat{m}_{WHM,g}(\vec{v})$ can also be thought
of as the reciprocal of an estimator of
$g(\vec{\theta})/(p(\vec{\theta})L(\vec{\theta}))$ using importance
sampling from the (unnormalized) posterior.  When using the estimator
$\widehat{m}_{WHM,g}(\vec{v})$, the denominator for each term in the
summation contains $p(\vec{\theta})L(\vec{\theta})$ evaluated at
points sampled from the posterior, so the limit to the variance of
this estimator will be the size (and quality) of the sample from the
posterior.  This seems acceptable, since these considerations will
limit any estimate of the marginal likelihood based on a sample from
the posterior.  Further, this seems more attractive than algorithms
which place the importance sampling density in the denominator, since
that could result in areas with sparse coverage (e.g., due to high
dimensionality) dominating the summation.  We show the performance of
$\widehat{m}_{WHM,g}(\vec{v})$ in Fig.\ 2, bottom, long-dashed curve.
While this estimator performs reasonably well, it has a larger
variance and appears to be converging more slowly than
$\widehat{m}_{IS,g}$ for our test case.  Additionally, We find that
this estimator is particularly sensitive to the choice of $\varsigma$
and rapidly degrades if $\varsigma$ is too large or too small.

\subsection{Sampling from Multiple Densities}
\subsubsection{Ratio Estimator}
We present a new estimator \citetext{Berger, private
communication}, based on the identity
\begin{equation}
m(\vec{v})= \int L(\vec{\theta})p(\vec{\theta}) d\vec{\theta} = \frac{\int p(\vec{\theta})L(\vec{\theta})h(\vec{\theta}) d\vec{\theta}} {\int h(\vec{\theta}) p(\vec{\theta} | \vec{v} ) d\vec{\theta}}\,.
\end{equation}
The key insight is to approximate the numerator by drawing a sample $\tilde{\theta}^1, \ldots
\tilde{\theta}^{n_s'}$ from $h(\vec{\theta})$ and to approximate the denominator
by drawing a sample  $\vec{\theta}^1, \ldots \vec{\theta}^{n_s}$ from the posterior (e.g., via MCMC).  This
yields the ratio of estimators,
\begin{equation}
\widehat{m}_{RE,h}(\vec{v}) = \frac{\frac{1}{n'_s} \sum_{i=1}^{n'_s} p(\tilde{\theta}^i)L(\tilde{\theta}^i)}
{\frac{1}{n_s} \sum_{i=1}^{n_s} h(\vec{\theta}^i)} \,.
\end{equation}
This estimator seems particularly promising, since both the numerator
and denominator are separate sums and there is no risk of a small
denominator leading to a large variance, as in importance sampling.
If we combine this estimator with the $g(\vec{\theta})$, the mixture
of normal distributions used in \S\ref{SecMixture} (again using a
distinct subsample from the posterior for constructing
$g(\vec{\theta})$), then we obtain the estimator $\widehat{m}_{RE,g}$.  
In our test case, the numerator converges significantly more rapidly
than the denominator, and so we choose $n'_s = n_s / 10$ with minimal
impact on the variance of the estimator.  This estimator performs very
well in our test case (Fig.\ 2, bottom, short-dashed curve).  It
converges as rapidly as any of the other estimators that we
considered, with the possible exception of importance sampling from a
single normal distribution.  Further $\widehat{m}_{RE,g}(\vec{v})$
appears to be quite robust, in that it does not display sudden jumps
when a single additional sample significantly changes the value of the
estimator, as is more common with most of the other estimators.
Unfortunately, we found that this estimator was less accurate on more
complex test cases, yet it showed no warning signs that the estimator
had not converged, even after $\sim10^7$ samples.

\subsubsection{Parallel Tempering}
\citet{Gregory05Book} introduced the method of parallel tempering for
estimating the marginal posterior probability for extrasolar planet
observations.  In parallel tempering, several Markov chains are run in
parallel, each with a slightly different target density,
$\pi_{\beta}(\vec{\theta}) = p(\vec{\theta}|\mathcal{M})
p(\vec{v}|\vec{\theta},\mathcal{M})^\beta$, where $\beta$ is an
inverse temperature parameter that varies between 0 and 1.  In the
parallel tempering algorithm, each Markov chain 
typically evolves according to the usual candidate transition PDFs,
but periodically the algorithm proposes an exchange of states between
two Markov chains that have slightly different values of $\beta$.  The
``high temperature'' Markov chains ($\beta\simeq0$) will explore a
very broad region in parameter space and can help the ``coldest''
Markov chain ($\beta=1$) to sample from the full posterior
distribution, even when there are narrow and widely separated peaks in
the posterior distribution \citep{Gregory05AAT}.  In principle, the
marginal posterior probability can also be calculated from the
ensemble of Markov chains, using
\begin{equation}
\widehat{m}_{PT}(\vec{v}) = p(\mathcal{M}) \exp \left\{
\int d\beta\, \sum_{i=1}^{n_s} \log\left[ p(\vec{v}|\vec{\theta}_{i,\beta},\mathcal{M}) \right] \right\},
\end{equation}
where $\vec{\theta}_{i,\beta}$ is the $i$th state in the Markov chain
with target distribution $\pi_{\beta}(\vec{\theta})$, and the integral
over $\beta$ is to be approximated by an appropriate weighted sum over
the values of $\beta$ used by the various Markov chains.  

We show the performance of $\widehat{m}_{PT}(\vec{v})$ using 32
tempering levels for the test data set in Fig.\ 3.  Based on five completely
independent realizations of the parallel tempering algorithm, we found
that four provided a reasonable estimate of the $m(\vec{v})$, but one
set of chains resulted in an estimate roughly twice as large as any
other estimator tested.  If this set of Markov chains had been the
only realization, none of the usual diagnostics would have recognized
that it had not yet converged.  Therefore, we feel that more work is
needed to understand the properties of marginal posterior estimates
obtained from parallel tempering. If the
sensitivity of the estimator were due to the slow convergence of the
highest temperature chains ($\beta\simeq0$), then their contribution
to the integral over $\beta$ could be approximated by an analytic
sampler from the prior (corresponding to $\beta=0$).  However, we have
verified that for our test case (see Table~\ref{TabErrorFactor}), the Markov chains with very small
values of $\beta$ make only minor contributions to the integral over
$\beta$. The second column of Table~\ref{TabErrorFactor} gives the 
fractional error that would result
if this decade of $\beta$ was not included and thus indicates the sensitivity of the result to that decade. Therefore, we suspect that the chains limiting the accuracy of $\widehat{m}_{PT}(\vec{\theta})$ have $\beta>10^{-3}$.

\begin{figure}[!ht]
\plotone{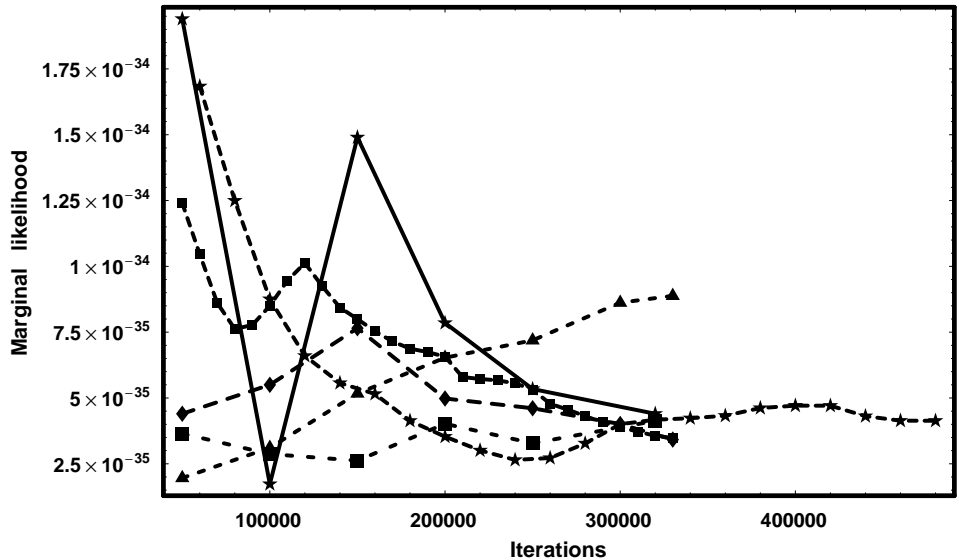}
\label{FigTemperingMarginalsVsSamples}
\caption{The estimate of the marginal probability for a one-planet
model and the 17 RV observations of HD88133 shown in Fig.\ 1 using the
parallel tempering method of \citet{Gregory05Book} as a function of
the number of iterations in each of the Markov chains.  Each of the
lines corresponds to a completely independent set of 32 Markov chains.
Since each curves is based on Markov chains with 32 different values
of $\beta$, even the parallel tempering simulations that stopped after
320,000 iterations required roughly five times more likelihood
evaluations than the each of the estimators shown in Fig.\ 2.}
\end{figure}

\begin{table}[!ht]
\caption{Fractional error versus $\beta$ for the results shown in Fig.\ 3.}
\smallskip
\begin{center}
\begin{tabular}{ll}
\tableline 
   $\beta$ range    &    Fractional error \\
\tableline
%
$1.0\ -\ 10^{-1}$      & $3.83 \times 10^{28}$  \\
$10^{-1}\ -\ 10^{-2}$  & $5.20 \times 10^{5}$  \\
$10^{-2}\ -\ 10^{-3}$  & $4.02$  \\
$10^{-3}\ -\ 10^{-4}$  & $0.54$  \\
$10^{-4}\ -\ 10^{-5}$  & $0.30$  \\
$10^{-5}\ -\ 10^{-6}$  & $0.26$  \\
$10^{-6}\ -\ 10^{-7}$  & $0.12$  \\
$10^{-7}\ -\ 10^{-8}$  & $0.02$  \\
$10^{-8}\ -\ 10^{-9}$  & $2 \times 10^{-3}$  \\
$10^{-9}\ -\ 10^{-10}$ & $2 \times 10^{-4}$  \\
$10^{-10}\ -\ 10^{-11}$ & $2 \times 10^{-5}$  \\
\tableline
\end{tabular}
\end{center}
\label{TabErrorFactor}
\end{table}

\subsection{Other Techniques for Model Selection}

While we are encouraged by the recent progress in developing efficient
and robust estimators of the marginal posterior probability, there are
several additional avenues of research that might lead to even more
desirable estimators.  One interesting method is based on the of
nested sampling methods of \citet{Skilling05}.  Unfortunately, for the
problem of radial velocity planet searches, there is no efficient way
to sample only from high posterior probability regions of parameter
space, as required by nested sampling.  As a result, we can only apply
nested sampling if we employ the very inefficient method of rejection
sampling.

Another class of algorithms for estimating Bayes factors relies on
sampling over the model space.  Two subclasses of methods have been
the subject of much research in the statistics community: reversible
jump MCMC \citep{Green95}, and model composition MCMC (MC$^3$)
\citep{CarlinChib95}.  Unfortunately, we find that the most obvious
choices of pseudopriors \citep[e.g.][]{GreenOHagan98} for MC$^3$
result in very poor mixing between different models.  Perhaps future
research can adapt these methods to allow for more rapid mixing
between models with different numbers of planets.  Similarly,
simplistic implementations of reversible jump methods seem unlikely to
be practical, since the trial jumps into higher dimensional spaces
will only land in areas of significant probability on extremely rare
occasions.  On the other hand, we are more optimistic about reversible
jump algorithms that employ an analytic approximation to each of the
posterior PDFs within the $i$th model
($p(\vec{\theta}|\vec{v},\mathcal{M}_i)$) for the transdimensional
steps.  We envision that each of the analytic approximations could be
based on mixture models constructed from a sample from
$p(\vec{\theta}|\vec{v},\mathcal{M}_i)$ obtained using conventional
MCMC techniques, similar to the importance sampling densities we
employed for $\widehat{m}_{IS,g}(\vec{v})$.

An even more radical idea is to abandon the computation of marginal
posterior probabilities in favor of some other statistic to aid in
quasi-Bayesian model selection.  Penalized likelihood methods such as
the Akaike and Bayesian information criterion do not seem well
justified when the posterior is significantly non-normal or
multi-modal.  Additionally, we are suspicious of any method that
penalizes all parameters equally, as our model is much more sensitive
to some model parameters (e.g., orbital period) than to others (e.g.,
eccentricity).  Therefore, we are more interested in exploring methods
based on the predictive distribution.  Unfortunately, any of these
alternative methods for model selection is somewhat arbitrary and less
than ideal for the purposes of adaptively scheduling observations based
on the principles of Bayesian adaptive experimental design.

\section{Conclusions}

In this paper, we have reviewed several methods for calculating the
marginal posterior probabilities in the context of RV planet searches.
One the positive side, we found that several algorithms were able to
accurately calculate the marginal posterior probability for a simple
test case, where there was a single dominant peak in the posterior
probability distribution.  However, all of the estimators based on
sampling from either the prior or posterior had serious short comings.

The method of partial linearization can be a useful tool for rapidly
computing relatively low accuracy estimates of $m(\vec{v})$ for data
sets with one planet or even $\sim1-3$ planets on low
eccentricity orbits.  However, it rapidly becomes computationally
intractable when there are multiple planets with significant
eccentricities.  Restricted Monte Carlo ($\widehat{m}_{RMC}(\vec{v})$)
can be useful for planets with large eccentricities, but is
computationally feasible only once the orbital parameters are
relatively well constrained.  Parallel tempering is able to estimate
$m(\vec{v})$ even for multimodal posterior distributions, but for our test data set
$\widehat{m}_{PT}(\vec{\theta})$ converged more slowly than all of the
other algorithms tested (except basic Monte Carlo).  For our test
case, we found no regime where the harmonic mean
($\widehat{m}_{NR}(\vec{v})$) or the weighted harmonic mean
($\widehat{m}_{WHM,h}(\vec{v})$) would be the most desirable
estimator.  The new ratio estimator ($\widehat{m}_{RE}(\vec{v})$)
performed very well for our test case, but we recommend proceeding
with caution, based on preliminary tests with more complex data sets.

Based on our tests, the most promising methods are based on importance
sampling using an analytic density that mimics the posterior (e.g.,
$\widehat{m}_{IS,N}(\vec{v})$ or
$\widehat{m}_{DIS,g^*,\vec{\beta}^*}(\vec{v})$).  When the posterior
has a single dominant peak that can be reasonably approximated by a
multivariate normal distribution, then simple importance sampling
($\widehat{m}_{IS,N}(\vec{v})$) provides a very efficient tool for
estimating marginal posteriors.  In cases where the posterior is more
complex (e.g., multiple peaks and/or non-linear parameter
correlations), then importance sampling can still be useful when
combined with a mixture distribution based on a sample from the
posterior that can be readily calculated via standard MCMC.
Refinements to the basic importance sampling algorithm (e.g.,
$\widehat{m}_{DIS,g^*,\vec{\beta}^*}(\vec{v})$) can provide increased
robustness and offer a tool for diagnosing when the mixture
distribution is sufficient.  We hope that future research will improve
our understanding of these estimator's theoretical and real-life
properties, as well as lead to additional refinements.  In particular,
we hope to investigate how the estimator
$\widehat{m}_{DIS,g^*,\vec{\beta}^*}(\vec{v})$ performs on more widely
dispersed posterior distributions and on higher dimensional problems
(e.g., multiple planet systems).

In a sense, we can consider the problem of Bayesian model selection to
have been reduced to the problem of constructing an analytic
approximation to a probability density based only on a set of samples
from the distribution.  Unfortunately, we recognize our algorithm for
constructing importance sampling densities is not yet
sufficiently robust to be applied generally.  Therefore, we would
like to see additional research that would improve the robustness and
computational efficiency of these algorithms.  Fortunately, we
recognize several ways our current algorithm could be improved.  For
example, rather than centering the mixture components on random
samples from the posterior, it might be possible to make do with a
smaller number of mixture components.  Perhaps methods making use of
Voronoi tessellations and/or quasi-Monte Carlo methods could be
beneficial in constructing good mixture distributions with fewer
components, improving the computational efficiency of these methods.

Finally, we note that this field is still young, and additional
research is needed to explore a wide range of methods for estimating
$m(\vec{\theta})$, including methods on importance sampling, parallel
tempering, reversible jump MCMC, MC$^3$, and nested sampling.

\acknowledgements 
The authors acknowledge many stimulating meetings of the exoplanets
working group during the Astrostatistics program at the Statistics and
Applied Mathematical Sciences Institute.  The working group was
supported in part by NSF grants AST-0507589 and AST-0507481.  The
authors especially thank Jim Berger, Floyd Bullard, Merlise Clyde, Tom
Loredo, and Bill Jefferys for their contributions.
E.B.F. acknowledges the support of the Miller Institute for Basic
Research.  Additional support for this work was provided by NASA
through Hubble Fellowship grant HST-HF-01195.01A awarded by the Space
Telescope Science Institute, which is operated by the Association of
Universities for Research in Astronomy, Inc., for NASA, under contract
NAS 5-26555. P.C.G. acknowledges the support of a grant from the
Canadian Natural Sciences and Engineering Research Council at the
University of British Columbia.

\end{document}